\documentclass[10pt,letterpaper]{article}
\usepackage{opex3}
\usepackage[draft]{hyperref}
\usepackage{amsmath}

\usepackage{graphicx}

\usepackage{ae} %%for Computer Modern fonts

\begin{document}

%%%%%%%%%%%%%%%%%% title page information %%%%%%%%%%%%%%%%%%
\title{Interscale Mixing Microscopy: \\numerically stable imaging of wavelength- scale objects with sub- wavelength resolution and far field measurements}

\author{Sandeep Inampudi$^1$, Nicholas Kuhta$^2$, and \\Viktor A. Podolskiy$^1*$}
\address{$^1$ Department of Physics and Applied Physics, University of Massachusetts at Lowell, One University Avenue,Lowell,MA,01854, USA}
\address{$^2$ Department of Physics, Oregon State University, 301 Weniger Hall, Corvallis, OR 97331, USA}
\email{viktor\_podolskiy@uml.edu} %% email address is required



\begin{abstract}
We present an imaging technique that allows the recovery of the transparency profile of wavelength-scale objects with deep subwavelength resolution based on far-field intensity measurements. The approach, interscale mixing microscopy (IMM), relies on diffractive element positioned in the near-field proximity to the object, to scatter information carried by evanescent waves into propagating part of the spectrum. A combination of numerical solutions of Maxwell equations and nonlinear fitting is then used to recover the information about the object based on far-field intensity measurements. The potential of the developed formalism to recover wavelength/20 features of wavelength-scale objects in presence of up to 10\% noise is demonstrated. 
\end{abstract}
%%%%%%%%%%%%%%%%%%%%%%% References %%%%%%%%%%%%%%%%%%%%%%%%%

%%%%%%%%%%%%%%%%%%%%%%%%%%  body  %%%%%%%%%%%%%%%%%%%%%%%%%%
\section{Introduction}
Numerous applications in materials, device characterization, security and biology require imaging with subwavelength resolution\cite{Dorad1,Dorad2,Dorad3,Dorad4}. However, the resolution limit of conventional optical microscopy is fundamentally limited by the diffraction limit to approximately half of the vacuum wavelength ($\lambda$) \cite{BornWolf}. Immersion microscopy\cite{SolImmMicr}, sparsity based computational microscopy \cite{SergevCompSensing} and several diffraction-based techniques\cite{SIM,SSIM,FSL,confocal}, discussed below, have been successful in improving the resolution of optical microscopy\cite{NarmanovCLEO2013}. However, resolution of realistic far-field imaging systems remains limited to approximately one-quarter of free-space wavelength. Label-free imaging with deep subwavelength resolution typically reserved to scanning near-field microscopy (SNOM)\cite{NSOM} or near-field tomography\cite{CarneyPRL,GovyadinovPRL} techniques that impose significant constraints on image acquisition rates and are strongly affected by the artifacts caused by tip-object interaction. Diffraction-imaging techniques capable of deep subwavelength resolution based on far-field field\cite{SentenacPRL} and intensity\cite{SukoInaging} measurements of have been recently proposed. In the latter approach, interscale mixing microscopy (IMM), the object is positioned in the near-field proximity to diffraction element that outcouples information about subwavelength features of the object to the far field where it can be detected by conventional means, followed by computational reconstruction of the object. In this work we present a computationally stable basis for analysis of such subwavelength information recovery and discuss the performance and limitations of the related numerical imaging. 

The optical radiation, scattered or emitted by the object, can be represented as a linear combination of plane waves\cite{BornWolf}. In such representation, the information about the features of the object with the typical feature size $\Delta$ is encoded in the plane waves having transverse wavevectors of the order of $k_\perp\sim 2\pi/\Delta$ \cite{GpAgarwalBook}. At the same time, evolution of this information in the direction along the imaging axis ($z$ direction in this manuscript) is given by $\propto\exp(i k_z z)$ with wavevector component $k_z$ related to $k_\perp$ and to angular frequency of optical radiation $\omega=2\pi c/\lambda$ via dispersion relation 
\begin{equation}
\label{dispEq}
k_\perp^2+k_z^2=n^2 \frac{\omega^2}{c^2}.
\end{equation}
It can be clearly seen that information corresponding to $\Delta\ll\lambda$ corresponds to imaginary values of $k_z$, and thus it exponentially decays away from the object\cite{NarmanovCLEO2013}. 

The main motivation behind improving the resolution of optical microscopy can be thus related to detection or reconstruction of the radiation corresponding to larger values of $k_\perp$. For some objects, recovery of evanescent part of the spectrum can be completed based on measurements of the propagation part of the spectrum\cite{SergevCompSensing}. Immersion microscopy-related techniques aim to increase effective refractive index $n$ and thus postpone the onset of diffraction limit. SNOM uses ultra-sharp tips to scatter (diffract) the radiation from the near-field of the object to the far-field zone; the shape of the tip is optimized to predominantly diffract information about small features of the object. Structured Illumination Microscopy (SIM)\cite{SIM} effectively doubles resolution by illuminating object with beams with $k_\perp\sim n\omega/c=k_0$ and analyzing the diffracted light. In a far-field superlens (FSL)\cite{FSL}, an object is first imaged directly, and is then imaged through a plasmonic resonant structure that is designed to amplify the information about subwavelength features of the object and scatter this information to the far-field by outcoupling it though first-order diffraction of the diffraction grating. In practical systems, resolution of both SIM and FSL is of the order of $\lambda/4$. 

In contrast to the above techniques, the IMM, originally proposed in Ref.\cite{SukoInaging}, is capable of performing imaging of wavelength-scale objects with deep subwavelength resolution. In the original proposal, the technique utilized a diffraction grating to break the translational symmetry and out-couple information corresponding to multiple length-scales of the object to the far-field via {\it multiple} diffraction orders of the grating. Hence, the resultant diffraction pattern measured at far-field contains  `mixed' contributions from both evanescent and propagating parts of the light scattered from the object. High resolution images are then reconstructed by using a series of computational post-processing steps aimed to `unmix' these contributions. While there is no fundamental limit on coupling between different parts of the spectrum, accuracy and resolution of the final image and the robustness of the technique are dependent on i) the efficiency and stabilty of the diffraction element to couple the spectrum, and ii) the efficiency of employed computational optimization techniques to un-couple the spectrum and cope with the noise in the measurement data\cite{NarmanovCLEO2013}. Therefore, both diffraction element and the numerical procedure can be optimized to improve image recovery.  

%Original proposal\cite{??} In the preliminary studies of this technique a simple subwavelength periodic grating has been used to couple the spectrum and a mathematical expansion of the object spectrum into polynomial basis has been used to uncouple the spectrum. Even though the demonstrated image resolutions are promising, the recovery mechanism is not stable enough to  span a distribution of subwavelength objects in larger areas.  The instability is primarily due to the mis-correlation between the used basis expansion to any physically relatable quantities of the system. 

In this work, we present highly stable wide-field image recoveries of wavelength-scale objects with resolutions of the order of $\lambda/20$ by representing the objects as collections of sub-wavelength ``pixels''. In addition, we propose new designs of diffraction elements that improve the resolution and stability of IMM technique as compared to simple periodic diffraction gratings.  

The rest of manuscript is organized as follows. In Section 2 we present details of the proposed formalism, and introduce the pixel basis. Section 3 presents analysis of imaging performance of periodic diffraction elements.  In Section 4 we analyze imaging performance of chirped diffractive elements. Section 5 concludes the manuscript. 

\section{Mathematical foundations of Interscale Mixing Microscopy (IMM)}
The schematic of the system is presented in Fig (1). The object is positioned in the near-field vicinity [at the distance $y_0\ll \lambda$] of the diffraction element (geometry of the diffraction element will be discussed below). The object is then excited by plane waves propagating at [multiple] incident angles $\theta_i$. Angular distribution of light intensity away from the diffraction element is recorded for each incident angle and is used to numerically reconstruct transparency of the object. 
\begin{figure}[htbp]
\centerline{\includegraphics[width=8cm]{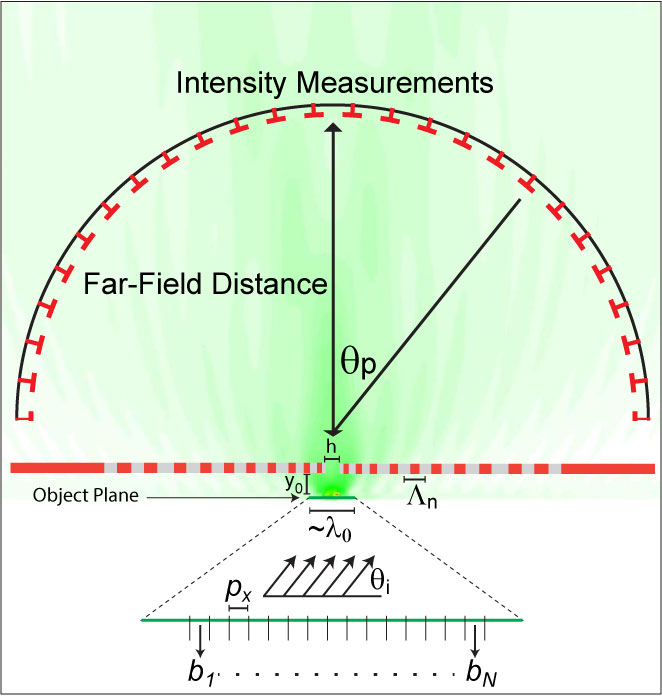}}
\caption{Schematic of IMM; Main figure: the diffraction-based high-resolution imaging setup; inset: line object represented as a set of pixels with unknown amplitudes $b_i$}
\end{figure}

In order to estimate the realistic performance of the presented technique, in this work we use ``computational experiment'' approach. In this approach, we use finite-element method (FEM) \cite{COMSOL} solutions of Maxwell equations to emulate experimental measurements and use rigorous-coupled-wave-analysis (RCWA) \cite{RCWA} solutions of Maxwell equations to perform image recoveries. The artifacts of FEM (mostly related to finite-simulation-space) are fundamentally different from artifacts of RCWA (related to implicit periodicity of the geometry); therefore, the pair of numerical techniques in some sense emulates real situation when data generated in experiment and image is recovered by a computer. As an added benefit, our technique allows us to compare recoveries based on intensity- and field-measurements (the latter are routinely performed at THz and GHz frequencies) and allows us to controllably add noise to our ``measurements''. In this work we restrict ourselves to recovering one dimensional (line-shaped) objects. Generalization of the presented approach to two-dimensional objects, although relatively straightforward (see, e.g. Ref.\cite{SentenacPRL} for example of field-based imaging), is devoted to future work. 

In the image reconstruction process, we begin by using RCWA to compute the grating-specific transfer function $\tau(k_x,\theta_p)$ that defines the contribution of information, originally encoded into plane waves with wavevector component $k_\perp=k_x$ at the object plane, to the far field propagating in the direction $\theta_p$. The transfer function is calculated for a broad spectrum of $k_x$ values spanning both the propagation $(|k_x|\le k_0)$ and evanescent $(|k_x|>k_0)$ regions. The far-field intensity of the light behind the diffraction element, emitted by an object in front of the element can be now related to the field that would be emitted by the object in the absence of the diffraction element. Explicitly, consider an isolated object excited by the field propagating along the $z$ axis. Assuming that the spectrum of the field scattered by this object is given by $a(k_x)$, the far-field intensity $I_p$ in the direction $\theta_p$ behind the diffraction element due to this same object excited by the plane wave coming from direction $\theta_i$ (see Fig.1) can be written as:
\begin{equation}
I_p(\theta_p,\theta_i) =\left|\sum_{k_x=-k_{max}}^{k_{max}}\tau(k_x,\theta_p)a(k_x-k_i)\right|^2,
\end{equation} 
where $k_i=k_0\sin(\theta_i)$ represents the wavevector component of the incident electromagnetic wave. 

The imaging problem is mathematically equivalent to recovering the unknown amplitude spectrum $a(k_x)$ based on measurements of $I_p$. Here we perform this task by minimizing the deviation between the measured intensity patterns $I_{meas}$ and calculated intensity patterns $I_p$ for multiple measurement directions $\theta_p$ and incident angles $\theta_i$, 
\begin{equation}
\sum_{\theta_i}\sum_{\theta_p}\left|I_p(\theta_p,\theta_i) -I_{meas}(\theta_p,\theta_i)\right|^2\to \rm min. 
\end{equation} 

It is critical to use sufficient number of $k_x$ terms in Eq.(2) in order to adequately calculate the distribution $I_p(\theta_p,\theta_i)$ and avoid the artifacts related to spectrum discretization. However, the limited number of measurements is typically not sufficient to solve for ``digitized'' version of the object spectrum $a(k_x)$. It is therefore required to develop a model that can parameterize the spectrum of the object with relatively few parameters, and use the optimization procedure [Eq.(3)] to deduce numerical values of these parameters. The original work \cite{SukoInaging} used Taylor series representation for the spectrum of the source. However, detailed analysis demonstrates that such model is overly sensitive to measurement noise, especially for recovering evanescent part of the spectrum $|k_x|\gg k_0$. Representing the spectrum as linear combination of orthogonal polynomials, as well as linear combination of harmonic- and Bessel- functions yields qualitatively similar results. Here we propose to represent the object as linear combination of finite-sized pixels and use Eq.(3) to calculate the amplitude of these pixels, essentially calculating the transparency profile of the object. On the implementation level, we divide the object plane into $N$ pixels; each pixel having width $p_x$ and centered around $x_n$ in the $y_0$ plane. Each such pixel produces electromagnetic field that is equivalent to a single slit diffraction pattern with an unknown amplitude $b_n$ with spectrum 
\begin{equation}
a(k_x-k_i)=\frac{\sin((k_x-k_i)p_x/2)}{2\pi(k_x-k_i)}\exp(ik_yy_0)\sum_{n=1}^Nb_n\exp[i(k_x-k_i)x_n]
\end{equation} 
Note that in principle $b_n$ can be extracted as a complex number that represents both amplitude and phase of the pixel. The main advantage of the pixel basis expansion is that the number of unknown quantities that need to be optimized now can be reduced depending on the object size and the target resolution. Most importantly, as seen below, these results are stable and have significant noise tolerance.

Two types of objects were studied in our ``numerical experiments''. The objects of the first type, ``sources'', represent the luminescent objects, represented in FEM as finite-sized line currents; these objects correspond to luminescent tags or finite-sized slits in the screen. The objects of the second type ``blocks'', are modeled as finite-sized perfect electric conductor lines, excited by incoming electromagnetic beams. In both cases, objects were positioned at $y_0 =\lambda/40$, and the field (intensity) was measured along the circular arc with radius of $30\lambda$; both measurement and incident angles $\theta_p, \theta_i$ spanned between  $-60^o$ and $60^o$, with increments of $1^o$ for $\theta_p$ and $20^o$ for $\theta_i$ respectively. In stability analysis, random noise $-1\le r_n\le 1$ is added to ``measured'' intensity:% (or field):
\begin{equation}
I_{meas}(\theta_p, \theta_i)\leftarrow |I_{meas}(\theta_p, \theta_i)+\delta\cdot r_n\cdot\max(I_{meas})|, 
\end{equation}
with parameter $\delta$ characterizing noise level. Nonlinear least square fit technique\cite{Fmincon} was used to perform optimization given by Eq.(3). In most recoveries, we aim to resolve $\lambda/20$ features of $\sim \lambda$-sized objects. 

\section{IMM with periodic diffraction elements}

We begin by analyzing the performance of the IMM technique with simplest possible diffraction element, periodic diffraction grating with period $\Lambda$. In order for the grating to provide substantial interscale mixing, and provide efficient coupling of evanescent information into propagating waves, grating period should be of the order of free-space wavelength. Gratings with substantially smaller periods couple evanescent waves only through high diffraction orders and loose efficiency. Gratings with substantially subwavelength period have Bloch vector $q=2\pi/\Lambda\gg k_0$, and thus leave ``gaps'' in the shifted spectrum. 
\begin{figure}[htbp]
\label{noiserecoveries}
\centerline{\includegraphics[width=13cm]{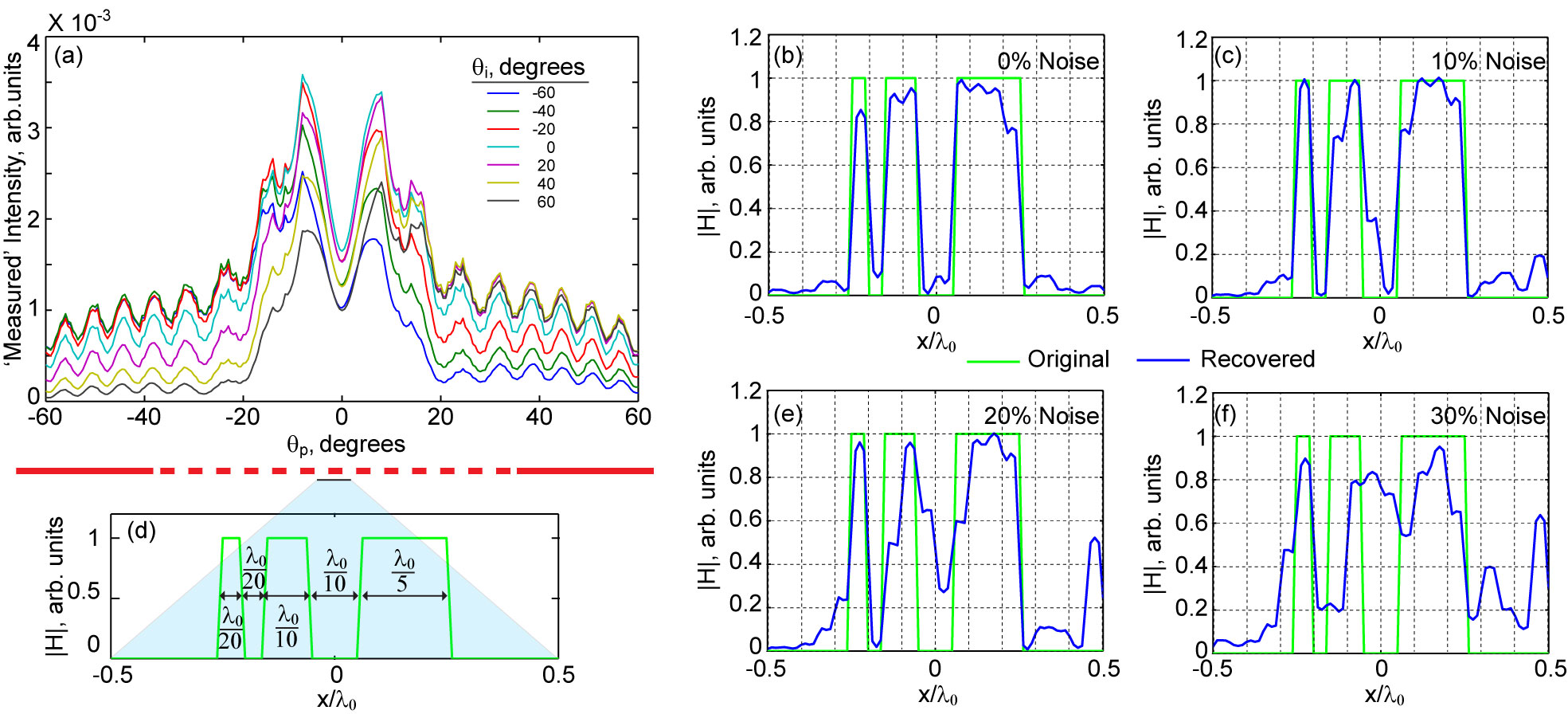}}
\caption{Image reconstruction using pixel basis expansion. (a) The far-field intensity pattern created by the array of subwavelength sources, schematically shown in (d). (b) Recovered image, based on intensity patterns shown in (a).  To demonstrate the stability of computational image recoveries, we added noise to the data in (a) and recovered the images starting from this noisy data. Panels (c), (e), and (f) represent recoveries corresponding to 10\%, 20\%, and  30\% noise respectively}
\end{figure}

%To demonstrate the stability of pixel basis expansion we designed `virtual experiments' using FEM software. One dimensional subwavelength sized objects with subwavelength separations are created to emit light at different angles. For simplicity a grating element with uniform periodicity $(0.7\lambda)$ and 60 periods is considered and placed close to the object at a distance of $\lambda/40$. To define the resolution independent of the immersion medium the objects and the grating are placed in free space. Far-field intensity is measured (calculated) at a distance of nearly $30\lambda$ from the grating with angular steps of $1$ degree from -60 to +60  degrees. Measurements are taken for 7 incident angles from -60 to +60 degrees in steps of 20 degrees.

Fig.\ref{noiserecoveries} illustrates imaging performance of periodic diffraction grating with $\Lambda=0.7 \lambda$ used to recover a set of three sources with sizes $\lambda/20$, $\lambda/10$, and $\lambda/5$. It is clearly seen that presented computational imaging technique is highly tolerant to [up to 20\% of] experimental noise. Interestingly, the noise-related artifacts in pixel basis do not always lead to disappearance of the smallest features of the composite source. Rather, the noise reduces the overall contrast level, and introduces parasitic sources. 

\begin{figure}[htbp]
\label{blocksRecovery}
\centerline{\includegraphics[width=13cm]{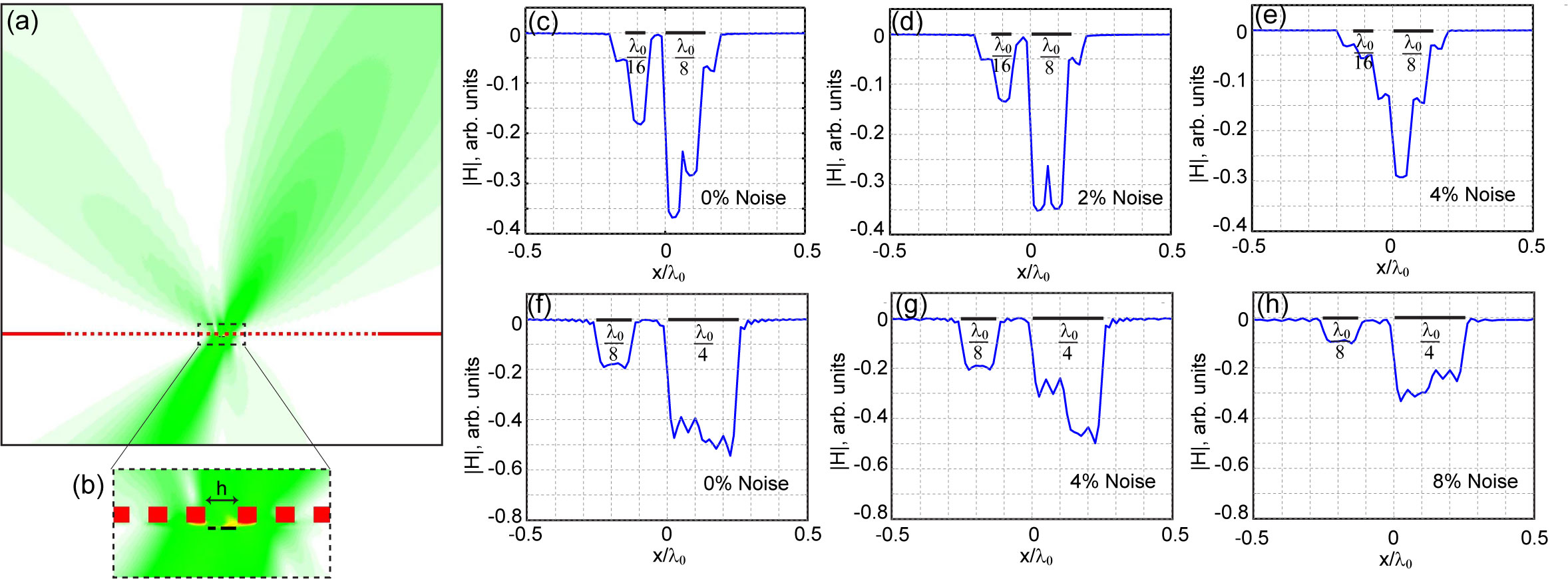}}
\caption{Image reconstruction of scattering objects (blocks) (a-b) The schematic of imaging system with external excitation of PEC line objects [black lines in (b)] using a Gaussian beam [color map].  (c\ldots e) Image recoveries objects that fit in the air gap $h$ of a compound grating with period of $0.5\lambda$. Panels (c), (d), and (e) correspond to noiseless recovery and to recovery with $2 \%$ and $4 \%$ random noise respectively. (f\ldots h) Image recoveries of objects with compound size greater than $h=0.55\lambda$; noiseless recovery is shown in (f), recoveries corresponding to 4\% and 8\% noise are shown in (g) and (h) respectively.}
\end{figure}

In the second case the block objects are illuminated with a gaussian beam with beam width of the order of few wavelengths in size as shown in Fig.\ref{blocksRecovery}(a). The measured intensity in the far-field now contains the contributions from both the input beam and the scattered light from the objects coupled through the grating. The recovery process is similar to the one described through Eqs.(2 \ldots 5). However, it becomes necessary to separate the contributions of the Gaussian background field that often dominates the measured intensity $I_p$. To perform such procedure, $I_p$  in Eq.(2) is now expressed as:    
\begin{equation}
I_p(\theta_p,\theta_i) =\left|H_g(\theta_p,\theta_i)+\sum_{k_x=-k_{max}}^{k_{max}}\tau(k_x,\theta_p)a(k_x-k_i)\right|^2
\end{equation} 
where $H_g(\theta_p,\theta_i)$ is the numerically calculated background field of the Gaussian beam propagating in the direction $\theta_i$, diffracted through the grating with no objects, and detected at an angle $\theta_p$. In experimental realizations, the field $H_g$ can be numerically calculated based on a set of calibration measurements of a fabricated grating.

As expected, presence of strong background field, along with substantial interaction between the objects and diffraction element, unfavorably affects the imaging performance of the proposed technique. However, despite these challenges, the proposed approach is substantially resilient to be able to recover objects of the order of $\lambda/16$ as shown in Fig.\ref{blocksRecovery} (b) with a noise tolerance of $2\%$ using a periodic diffraction grating having a period of $0.5 \lambda$. Note that since the noise level is defined as percentage of measured intensity, the noise introduced into the measurements is comparable (and even sometimes bigger than) the perturbation of far-field intensity caused by scattering of light by the objects. 

During the image recovery process, specifically in the case of blocks, it is seen that the objects residing behind the air gaps of the diffraction grating are recovered more accurately than the objects placed behind the metallic parts of the grating. Therefore, to image larger objects, we propose to use a diffraction element where the two grating arms are separated by a gap with pre-selected size $h(\approx\lambda)$. Such compound-grating system minimizes the multiple reflections between objects and the metallic parts of the grating and enables accurate recoveries of relatively large objects as illustrated in Fig.\ref{blocksRecovery}~(e-g). Another advantage of the compound-grating system comes from the fact that such design introduces a natural focal point into the imaging system. Furthermore, it can be straightforwardly extended to 2D scanning microscopes, potentially resulting in highly parallel SNOMs that image $\sim \lambda\times\lambda$-regions with deep subwavelength resolution.

\section{Image reconstruction using Aperiodic Diffraction Elements. }
Naturally, periodic diffraction gratings are just one (and possibly, simplest) example of diffraction elements. Numerous diffraction elements have been recently proposed for controlling light refraction and for generation of unconventional optical beams\cite{hasmanGrating,oldGrating,capassoGrating,shalaevGrating}. Similar meta-surfaces can be designed to increase the mixing efficiency of a diffraction element used for proposed here computational imaging. In this work, we restrict ourselves to analyzing the perspectives offered by aperiodic ``chirped'' gratings with linearly varying periodicity in both directions form center. 
\begin{figure}[htbp]
\label{Chirped}
\centerline{\includegraphics[width=12cm]{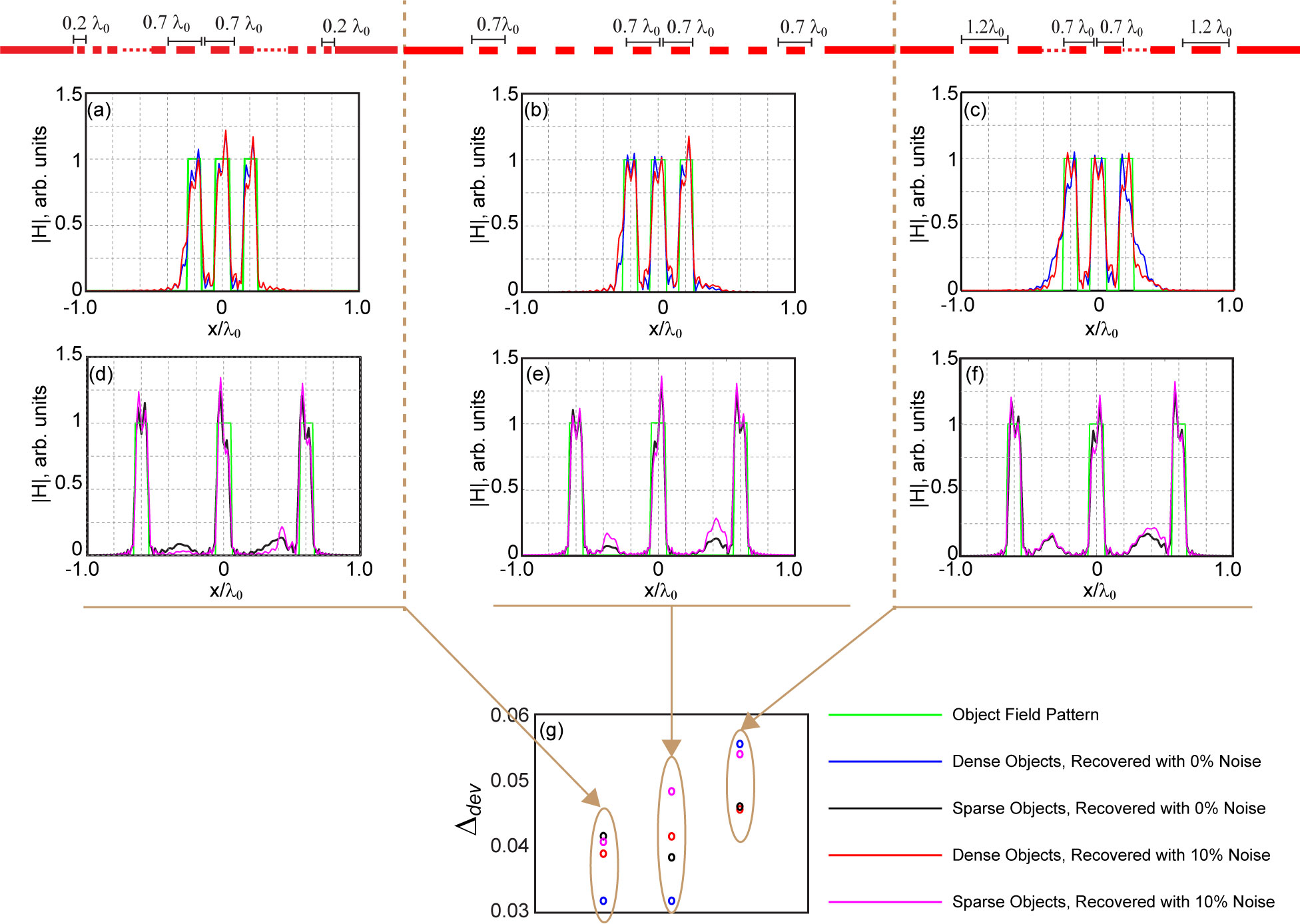}}
\caption{Image reconstruction using Aperiodic Diffraction Elements. Image recoveries of dense object (a)-(c) [$\lambda/10$ sized sources with $\lambda/10$ spacing]and sparse object (d)-(f) [$\lambda/10$ sized sources with $\lambda/2$ spacing] using three types of gratings with decreasing periodicity from center to ends (a,d), equal periodicity from center to ends (b,e) and increasing periodicity from center to ends (c,f). While all grating systems are able to recover all three objects, the system with decreasing periodicity tends to minimize the appearance of parasitic objects in the case of noisy measurements. Panel (g) quantifies the quality of image recovery via mean square deviation between recovered and original fields.}
\end{figure}
Three grating designs are analyzed. Periodicity of the first grating varies between $0.7\lambda$ at the center to $0.2\lambda$ at the ends with 30 elements on each side from the center; periodicity of the second grating is unchanged at $0.7\lambda$; periodicity of the last grating increases from $0.7\lambda$ to $1.2\lambda$ with 30 elements on each side from the center. Image recovery with all three designs of diffractive elements is tested for noise tolerance with sparse and dense objects (Fig.4). In contrast with a fully periodic grating, linearly changing periodicity in the two arms of the grating creates a well-defined focal region in the central position of the imaging system. Chirped grating also smooth-out diffraction performance of the element, distributing the mixed information over the measurement space. It is seen that a diffraction element with decreasing periodicity from center to the two sides of the arms performs better in imaging both dense and sparse objects in the presence of noise when compared to the other two types of elements tested. The grating element with decreasing periodicity is also relatively immune to the appearance of the parasitic objects, particularly in the case of recovering sparse objects from noisy measurements. 

To quantify performance of the particular diffraction element, we calculated mean square deviation between original and recovered fields:
\begin{equation}
\Delta_{dev}=\frac{1}{N}\sum_{n=1}^N \left|H_e(x_n)-H_r(x_n)\right|^2
\end{equation} 
where $H_e(x_n)$ is the field excited at pixel position $x_n$ and $H_r(x_n)$ is the field recovered at the pixel position $x_n$. It is clearly seen that the periodic system is superior for noiseless recovery (the fact that is possibly related to the underlying RCWA technique used in our numerical recoveries). At the same time, when recovering objects with noisy measurements, chirped grating with decreasing periodicity outperforms its counterparts. 

\section{Conclusion}
To conclude, we presented a computational imaging technique based on diffraction measurements. The presented technique allows to image wavelength-scale objects with deep subwavelength resolution. Convenient parameterization of objects was developed. Stability of the numerical image recoveries as a function of object size, separation, and diffraction element design was analyzed. 

This research is supported by NSF (Grant \# ECCS-1102183)
\end{document}